\documentclass[aps,pra,onecolumn,superscriptaddress,nofootinbib]{revtex4-2}

\usepackage{amsmath, empheq,braket,amsthm,amssymb,graphics,amsfonts,array,color,subfigure, mathrsfs, dsfont, physics, float}
\usepackage{MnSymbol}

\newcommand*{\bigchi}{\mbox{\Large$\chi$}}

\definecolor{myurlcolor}{rgb}{0,0,0.7}
\definecolor{myurlcolor1}{rgb}{0,0.7,0.1}
\definecolor{myrefcolor}{rgb}{0,0,0.7}
\usepackage[unicode=true,pdfusetitle, bookmarks=true,bookmarksnumbered=false,bookmarksopen=false, breaklinks=false,pdfborder={0 0 0},backref=false,colorlinks=true, linkcolor=myrefcolor,citecolor=myurlcolor1,urlcolor=myurlcolor]{hyperref}

\usepackage{tensor} 
\usepackage{scalerel} 

\usepackage{url} 

\newcommand{\kb}[2]{\left| #1 \vphantom{#2} \right>\left< #2 \vphantom{#1} \right|} 
\newcommand{\proj}[1]{\kb{#1}{#1}} 

\DeclarePairedDelimiterX{\infdivx}[2]{(}{)}{%
  #1\;\delimsize\|\;#2%
}

\newtheorem{lem}{Lemma}

\newtheorem{conj}{Conjecture}

\newcommand {\info } {\! : \!}

\begin{document}

\title{Classical capacity of quantum non-Gaussian attenuator and amplifier channels\footnote{Contribution to IJQI special issue for A. S. Holevo's 80th birthday.}}

\author{Zacharie Van Herstraeten}
\affiliation{James C. Wyant College of Optical Sciences, University of Arizona, Tucson, AZ 85721, USA}

\author{Saikat Guha}
\affiliation{James C. Wyant College of Optical Sciences, University of Arizona, Tucson, AZ 85721, USA}

\author{Nicolas J. Cerf}
\affiliation{James C. Wyant College of Optical Sciences, University of Arizona, Tucson, AZ 85721, USA}
\affiliation{Centre for Quantum Information and Communication, \'{E}cole polytechnique de Bruxelles,  CP 165, Universit\'{e} libre de Bruxelles, 1050 Brussels, Belgium}

\begin{abstract}
We consider a quantum bosonic channel that couples the input mode via a beam splitter or two-mode squeezer to an environmental mode that is prepared in an arbitrary state. We investigate the classical capacity of this channel, which we call a non-Gaussian attenuator or amplifier channel. If the environment state is thermal, we of course recover a Gaussian phase-covariant channel whose classical capacity is well known. Otherwise, we derive both a lower and an upper bound to the classical capacity of the channel, drawing inspiration from the classical treatment of the capacity of non-Gaussian additive-noise channels. We show that the lower bound to the capacity is always achievable and give examples where the non-Gaussianity of the channel can be exploited so that the communication rate beats the capacity of the Gaussian-equivalent channel (\textit{i.e.}, the channel where the environment state is replaced by a Gaussian state with the same covariance matrix). Finally, our upper bound leads us to formulate and investigate conjectures on the input state that minimizes the output entropy of non-Gaussian attenuator or amplifier channels. Solving these conjectures would be a main step towards accessing the capacity of a large class of non-Gaussian bosonic channels.
\end{abstract}

\maketitle

\section{Introduction}

\begin{figure}[t]
\centering
\includegraphics[width=0.7\linewidth]{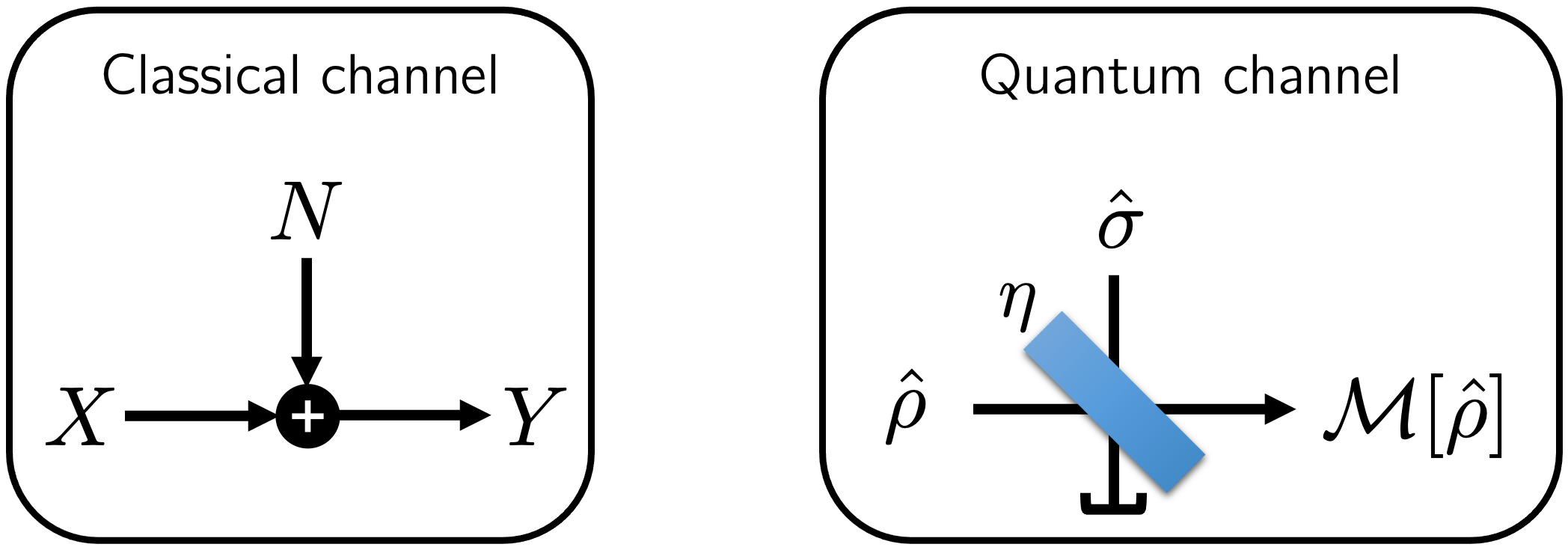}
\caption{
Left panel : Classical non-Gaussian additive-noise channel, where $X$, $Y$, and $N$ denote respectively the input, output, and noise random variables. The noise $N$ admits an arbitrary probability density.  Right panel : quantum non-Gaussian attenuator channel. The input state $\hat \rho$ is coupled with an arbitrary environment state $\hat \sigma$ through a beam splitter of transmittance $\eta$, resulting in an output state ${\cal M}[\hat \rho]$ while tracing out the other output mode. Here ${\cal M}$ denotes the corresponding trace-preserving completely-positive map acting on $\hat \rho$.
}
\label{fig:classical_quantum_channel}
\end{figure}

It only took a few decades before Shannon's groundbreaking mathematical theory of information \cite{Shannon1948-ty} was propelled to a whole new level through the advance of quantum mechanics.
With the field of quantum information, the theoretical framework describing the transmission of information over any physical medium was born.
In particular, it provided a precise characterization of the optimal rate of data transmission over a quantum channel, defined as the \textit{classical capacity} or \textit{quantum capacity} depending on the nature of the data being transmitted.

When it comes to classical data, the information content retrievable from a quantum state is subject to a fundamental upper bound known as the \textit{Holevo information} \cite{Holevo1973-yo}, which caps the amount of classical bits extractable from the system.
The classical capacity of a channel is then obtained by maximizing the Holevo information at the output of the channel over every possible input ensembles \cite{PhysRevA.56.131,651037}, usually subject to an energy constraint in the case of infinite-dimensional systems.
Throughout the present paper, our focus will be on the classical capacity and we will sometimes simply refer to it as the capacity of the channel.

Over the years, there has been a continuous effort towards studying the classical (and quantum) capacities of various types of quantum channels, leading in some cases to exact solutions.
In this endeavor, the family of bosonic Gaussian channels (BGCs) has attracted a particular interest \cite{Holevo2001-qb} as BGCs adequately model most optical links of modern communication systems.
Their classical capacity was exactly solved in Refs. \cite{Giovannetti2014} for the most common case of phase-insensitive (and phase-conjugating) channels, based on the solution of a Gaussian optimizer conjecture \cite{Giovannetti2015-mz}. 
More general cases of phase-sensitive BGCs were studied in Refs. \cite{PhysRevLett.111.030503,schafer2016}.
Besides Gaussian channels, let us mention the depolarizing channel, whose capacity was solved in Ref. \cite{King2003-ra}, and, on a more recent note, the amplitude damping and dephasing channels, whose minimum output entropy was studied in Ref. \cite{Memarzadeh2016-cc}.
The quantum capacity of the dephasing channel was studied in Refs. \cite{Arqand2020-po, Dehdashti2022-fj} and solved in Ref. \cite{Lami2023-jr}.

It is only recently that the communication over non-Gaussian quantum channels has started to attract attention. 
For example, the quantum capacity of so-called general attenuators was considered in Refs. \cite{Lim2019-uf, PhysRevLett.125.110504}.
General attenuators couple an input state with an arbitrary environment through the action of a beam-splitter, and can be understood as the quantum analog of the (scaled) addition of two random variables, as depicted in Fig. \ref{fig:classical_quantum_channel}.
From a more fundamental perspective, several entropic properties of quantum attenuators have been studied and conjectured in Refs. \cite{Guha2007, Guha2016-xe}.
Note that the case of a general attenuator with a Fock environment belongs to the family of photon-added Gaussian channels as defined in Ref. \cite{PhysRevA.95.062309}.

Here, we will consider a specific type of non-Gaussian quantum channels whose Stinespring dilation involves a Gaussian unitary, specifically a beam-splitter or two-mode squeezer. 
If the environment is initially in a thermal state, the channel reduces to the well-known (Gaussian phase-insensitive) quantum attenuator or amplifier channel, whose classical capacity was derived in Ref. \cite{Giovannetti2014}. 
Instead, we will assume here that the environment state is arbitrary, which yields what we call a non-Gaussian attenuator or amplifier channel. We will be interested in the ultimate classical communication rate via this channel, namely its classical capacity (we actually restrict ourselves to the single-shot capacity, ignoring the potential complication brought by superadditivity \cite{Hastings2009-cs}).

Our main result is that we are able to constrain the classical capacity of such a non-Gaussian channel $C({\cal M})$ by using the known capacity of the corresponding Gaussian-equivalent channel $C({\cal M}_G)$, namely 
\begin{equation}
\label{eq:main-result}
C({\cal M}_G)\le C({\cal M}) \le C({\cal M}_G) +\Delta ,
\end{equation}
where the width $\Delta$ of this interval is linked to the non-Gaussian character of the channel ${\cal M}$ (it can be explicitly expressed in some cases, or numerically computed).
In addition, we conduct numerical exploration to investigate the minimum output entropy (MOE) of various quantum non-Gaussian attenuators.  Little is known about this quantity as of today although it is a key ingredient in order to compute the width $\Delta$, that is, the potential capacity increase due to non-Gaussianity.

This paper is organized as follows. We first recall in Sec. \ref{sec:classical} the treatment of the capacity of classical non-Gaussian additive-noise channels in order to stress the similarities and discrepancies between the classical and quantum scenarios. Then, in Sec. \ref{Sec:quantum}, we derive lower and upper bounds on the classical capacity of quantum non-Gaussian attenuator or amplifier channels. This leads us to investigate the width $\Delta$ in Sec. \ref{sec:conjectures}
 and, in particular, formulate conjectures on the minimum output entropy of certain classes of quantum non-Gaussian channels (especially when the environment state is Fock-diagonal or admits some discrete symmetries). Finally, we conclude in Sec. \ref{sec:conclusions}.

\section{Classical additive-noise channel}
\label{sec:classical}

Consider an arbitrary (non-Gaussian) additive-noise channel (see Fig.~\ref{fig:classical_quantum_channel}) that is modeled by 
\begin{equation}
X \to Y = X+N ,
\end{equation}
where $X$ denotes the input, $Y$ denotes the output, and $N$ denotes the noise. Here, $X$, $Y$, and $N$ are real-valued random variables and $N$ is assumed to be independent of $X$, which implies that $\sigma^2_Y = \sigma^2_X +\sigma^2_N$. We will also use the quantum-inspired notation
\begin{equation}
{\cal M}: \quad p_X \to p_Y ={\cal M}[p_X] ,
\end{equation}
where $p_X$ and $p_Y$ are the probability densities of the input and output variables, and the channel is viewed as a stochastic map $\cal M$ depending on the probability density $p_N$ of the noise. Without loss of generality, we restrict ourselves  to probability densities of the input $p_X$ and noise $p_N$ that have a zero mean (hence, the same is true for $p_Y$). We will follow Ref. \cite{IHARA197834} to express lower and upper bounds on the capacity of $\cal M$, which is defined as
\begin{equation}
\label{eq:def-classical-capacity}
C ({\cal M}) \coloneqq \max_{p_X\in {{\mathcal P}_E}} I(X \info X+N) ,
\end{equation}
where $I(X \info Y)=h(Y)-h(Y|X)$ stands for the mutual information between variables $X$ and $Y$, with $h(\cdot)$ standing for Shannon differential entropy and ${{\mathcal P}_E}$ standing for the set of probability densities with a variance that does not exceed $E$ (interpreted as an energy), namely
\begin{equation}
{\mathcal P}_E \coloneqq \left\{ p_X ~:~ \sigma^2_X = \int \! x^2 \, p_X(x) \, \mathrm dx \le E \right\} .
\end{equation}
Since $I(X \info X+N)=h(X+N)-h(X+N|X)=h(X+N)-h(N)$ as the noise is independent of the input, the capacity may be rewritten as 
\begin{equation}
C ({\cal M})= \max_{p_X\in {{\mathcal P}_E}}  h(X+N) - h(N) ,
\end{equation}
where the maximization does not involve the second term as it only depends on the noise\footnote{This is in sharp contrast with the classical capacity of quantum channels, where the corresponding negative term depends both on the noise and on the input, see Sec. \ref{Sec:quantum}.}.

\subsection{Capacity of the Gaussian-equivalent channel}

We define the Gaussian-equivalent channel as the map ${\cal M}_G$ obtained when the noise $N$ is replaced by a Gaussian noise $N_G$ with the same variance ($\sigma^2_N=\sigma^2_{N_G}$), namely
\begin{equation}
X \to Y = X+N_G  ,
\end{equation}
or, using the quantum-inspired notation, 
\begin{equation}
{\cal M}_G : \quad p_X \to p_Y ={\cal M}_G[p_X]  .
\end{equation}
Using Shannon theory, the capacity of this Gaussian channel can be expressed exactly because the maximum in Eq. \eqref{eq:def-classical-capacity} is attained when the input is Gaussian distributed with variance $E$, so when $X$ is replaced by $X_G$ (with $\sigma^2_X=\sigma^2_{X_G}$). Hence, we have
\begin{eqnarray}
C ({\cal M}_G)&=& I(X_G \info X_G+N_G) \nonumber \\
&=& h(X_G+N_G) - h(N_G) .
\end{eqnarray}
Since both $X_G+N_G$ and $N_G$ are Gaussian-distributed, their entropy can be written explicitly as
\begin{eqnarray}
h(X_G+N_G) &=& {1 \over 2} \ln (2\pi e \, (\sigma^2_{X_G} + \sigma^2_{N_G} )) , \\
h(N_G)  &=& {1 \over 2} \ln (2\pi e \, \sigma^2_{N_G}) ,
\end{eqnarray}
which yields the standard expression
\begin{eqnarray}
C ({\cal M}_G)&=& {1 \over 2} \ln(1+\gamma) ,
\end{eqnarray}
with $\gamma = \sigma^2_{X_G} / \sigma^2_{N_G}$ being the signal-to-noise ratio. 
Starting from $C ({\cal M}_G)$, we will now prove lower and upper bounds on $C({\cal M})$, namely Eq. \eqref{eq:main-result} where the width will be expressed as a relative entropy, \textit{i.e.}, $\Delta_\text{cl}= D(N || N_G)$, which measures the non-Gaussianity of the noise $N$.

Let us indeed recall the fact that the relative entropy between any random variable $X$ and the corresponding Gaussian-distributed random variable $X_G$ is given by 
\begin{equation}
D(X||X_G) = h(X_G) - h(X) \ge 0 ,
\label{eq:relative-entropy}
\end{equation}
where this expression as an entropy difference holds provided $X$ and $X_G$ have the same mean and variance (recall that the mean of all variables is set to zero here). Inequality \eqref{eq:relative-entropy}
expresses the fact that the Gaussian probability density has the highest entropy among all probability densities with the same variance. Hence the difference $D(X||X_G)$ is a measure of the non-Gaussianity of $X$.

\subsection{Lower bound on $C ({\cal M})$}

Among all distributions with the same variance, the optimal distribution $p_X$ that achieves the capacity $C ({\cal M})$ of the arbitrary channel ${\cal M}$ has no reason to be the Gaussian distribution $X_G$ that achieves the capacity of the Gaussian-equivalent channel ${\cal M}_G$. Hence, we get a lower bound on $C ({\cal M})$ by injecting the Gaussian input $X_G$ into channel ${\cal M}$, namely 
\begin{equation}
C ({\cal M})\ge  I(X_G \info X_G+N) .
\label{eq-lower-bound-C}
\end{equation}
We now compare the situations where this same Gaussian input $X_G$ is injected into the channel ${\cal M}$ or into the Gaussian-equivalent channel ${\cal M}_G$. We have
\begin{eqnarray}
\lefteqn {I(X_G \info X_G+N) - I(X_G \info X_G+N_G) } \nonumber \\
&=&[h(X_G+N)-h(N)]-[h(X_G+N_G)-h(N_G)] \nonumber \\
&=&[h(N_G)-h(N)]-[h(X_G+N_G)-h(X_G+N)] \nonumber \\
&=& D(N || N_G) - D(X_G+N || X_G+N_G) \nonumber \\
&\ge& 0 ,
\label{eq:ineq-lower-bound}
\end{eqnarray}
where we may write relative entropies because $N$ and $N_G$ (resp. $X_G+N$ and $X_G+N_G$) have the same variance. The inequality comes from the data-processing inequality for the relative entropy $D(N || N_G)$, when variables $N$ and $N_G$ are both processed in a channel with additive Gaussian noise $X_G$. Finally, combining inequalities \eqref{eq-lower-bound-C} and  \eqref{eq:ineq-lower-bound}, we have
\begin{equation}
C ({\cal M})\ge I(X_G \info X_G+N) \ge C ({\cal M}_G) ,
\end{equation}
where $C ({\cal M}_G) = I(X_G \info X_G+N_G) $ is indeed the capacity of ${\cal M}_G$ since it is achieved by $X_G$. The capacity of the Gaussian-equivalent channel $C ({\cal M}_G)$ is thus a lower bound on $C ({\cal M})$. Note that this lower bound is achievable since injecting the Gaussian input $X_G$ into channel ${\cal M}$ gives the rate $I(X_G \info X_G+N)$, which cannot be lower than the lower bound $C ({\cal M}_G)$.

The interpretation is straightforward. The Gaussian noise $N_G$ can only have a larger entropy than $N$ since $\sigma^2_N=\sigma^2_{N_G}$, so it is somehow more detrimental to the transmission of information than $N$. Thus, replacing the noise $N$ by its Gaussian equivalent $N_G$ makes a Gaussian channel that can only transmit less information, hence its lower capacity.

\subsection{Upper bound on $C ({\cal M})$}

We now express the difference between the mutual information when an arbitrary input $X$ is injected in the channel with arbitrary noise $N$ and when the input $X_G$ (with variance $\sigma^2_{X_G}=\sigma^2_X$) is injected in the Gaussian channel (with noise variance $\sigma^2_{N_G}=\sigma^2_N$):
\begin{eqnarray}
\label{eq-classical-upper-bound}
\lefteqn {I(X \info X+N) - I(X_G \info X_G+N_G) } \nonumber \\
&=&[h(X+N)-h(N)]-[h(X_G+N_G)-h(N_G)] \nonumber \\
&=&[h(N_G)-h(N)]-[h(X_G+N_G)-h(X+N)] \nonumber \\
&=& D(N || N_G) - D(X+N || X_G+N_G) \nonumber \\
&\le& D(N || N_G) ,
\end{eqnarray}
where we get relative entropies because $N$ and $N_G$ (resp. 
$X+N$ and $X_G+N_G$) have the same variance. The inequality in Eq. \eqref{eq-classical-upper-bound} comes from $D(X+N || X_G+N_G)\ge 0$, so that we have
\begin{equation}
I(X \info X+N) \le  I(X_G \info X_G+N_G) + D(N || N_G).
\label{eq:precurs-class-upp-bound}
\end{equation}

We must maximize $I(X \info X+N)$ over $p_X\in {{\mathcal P}_E}$ in order to find $C ({\cal M})$, but the right-hand side of this inequality does not directly depend on $X$ (it only depends on its variance via $X_G$). Hence,  we obtain
\begin{equation}
C ({\cal M})\le C ({\cal M}_G) + D(N || N_G) .
\end{equation}
Thus, for a classical non-Gaussian channel, the width of the interval is given by $\Delta_\text{cl} = D(N || N_G)$, which is the relative entropy between the noise $N$ and the Gaussian-equivalent noise $N_G$, measuring the non-Gaussianity of the additive noise of the channel. The more non-Gaussian is the noise, the larger is the potential capacity increase with respect to the Gaussian-equivalent channel. This capacity increase is nevertheless restricted to $\Delta_\text{cl} $ at most, which tends to zero when the noise $N$ tends to a Gaussian noise $N_G$, in which case both lower and upper bounds converge to 
$C ({\cal M}_G)$ as expected.

\section{General quantum attenuator or amplifier channels}
\label{Sec:quantum}

We will now transpose the above scenario to non-Gaussian quantum channels. We will first consider the general (non-Gaussian) quantum attenuator map shown in Fig.~\ref{fig:classical_quantum_channel}:
\begin{equation}
{\cal M}:\quad \hat\rho \to {\cal M}[\hat\rho] = \hat\rho \, \boxplus_\eta  \, \hat\sigma \coloneqq \Tr_2[\hat{U}_\eta(\hat\rho\otimes\hat\sigma)\hat{U}^\dagger_\eta]  .
\end{equation}
It is realized with a beam splitter (denoted with $\boxplus$) of arbitrary transmittance $\eta$ ($0\le \eta \le 1$) that couples the input state $\hat \rho$ with an environment prepared in an arbitrary state $\hat \sigma$ (we may assume with no loss of generality that the latter has a zero mean vector). We also assume the input and environment to be initially in a product state (just like the classical noise is independent of the input), so that the covariance matrix of the output state can be expressed as
\begin{equation}
\mathrm{Cov}\left({\cal M}[\hat{\rho}]\right) =  
\eta \, \mathrm{Cov}\left(\hat{\rho}\right) + (1-\eta) \, \mathrm{Cov}\left(\hat{\sigma}\right) .
\end{equation}

Further, we may assume with no loss of generality that the environment state has a covariance matrix proportional to the identity, $\mathrm{Cov}\left(\hat{\sigma}\right)\propto\mathbf{I}_2$. Indeed, if it is not the case, we may apply a squeezing unitary $U$ to $\hat \sigma$ so that $\mathrm{Cov}\left(U\hat{\sigma}U^\dagger \right)\propto\mathbf{I}_2$. 
Then, by applying the same $U$ on the input state $\hat\rho$, we can move $U$ to the output state, namely
$(U \hat\rho U^\dagger) \, \boxplus_\eta  \, (U \hat\sigma U^\dagger) = U (\hat\rho \, \boxplus_\eta  \, \hat\sigma) U^\dagger$, see Sect. \ref{subsec:prop_quantum_attenuator} and Fig. \ref{fig:beamsplitter_commutation}. 
Since acting with $U$ at the output of the channel is reversible, the capacity of the channel with modified environment is simply equal to $C({\cal M})$ provided each input state $\rho$ is replaced by $U\rho U^\dagger$.\footnote{Since the channel capacity is defined with an energy constraint on the average input state, one should actually take into account the energy cost of this extra squeezing $U$. 
Thus, the freedom to convert the environment to a state whose covariance matrix is proportional to the identity is only valid above some energy threshold \cite{PhysRevA.80.062313,PhysRevA.84.032318}. 
We restrict to this regime in the present analysis.} 
Thus, we will restrict to quantum attenuator maps whose environment state has a zero mean and a covariance matrix proportional to the identity. Since the treatment of the quantum amplifier map is very similar, we will only sketch it in Sec.~\ref{sec:quantum-amplifier-channel}.


For a quantum channel, the mutual information must be replaced by the classical-quantum mutual information (or Holevo information\cite{Holevo1973-yo}), namely, 
\begin{equation}
\bigchi \left[ \{p_i,\hat \psi_i\},{\cal M}\right] = 
S\left({\cal M}[\sumint_i p_i \hat \psi_i]\right)-\sumint_i p_i S\left({\cal M}[\hat \psi_i]\right)  ,
\end{equation}
which is a function of the input ensemble $\{p_i,\hat \psi_i\}$ and of the map ${\cal M}$. Here, $S(\cdot)$ stands for the von~Neumann entropy and we note that the ensemble can very well be realized by a continuum of input states, in which case the sum over $i$ would be replaced by an integral (hence, the hybrid notation). In analogy with Eq. \eqref{eq:def-classical-capacity}, the classical capacity\footnote{We focus on the single-shot capacity and will not consider the regularization problem here, namely the fact that the ultimate capacity of ${\cal M}$ should be written as $\lim_{n\to\infty} C({\cal M}^{\otimes n})/n \stackrel{?}{=} C({\cal M})$ .}  of ${\cal M}$ is expressed as a maximum
 \begin{equation} \label{eq:maximum_holevo_information}
C ({\cal M}) \coloneqq\max_{\displaystyle \{p_i,\hat \psi_i\} \in {\cal E}_\nu} \bigchi \left[\{p_i,\hat \psi_i\}, {\cal M} \right] 
\end{equation}
where ${\cal E}_\nu$ is the set of all ensembles $\{p_i,\hat \psi_i\}$ whose average photon number is at most $\nu$, namely 
\begin{equation}
{\cal E}_\nu \coloneqq \left\{ \{p_i,\hat \psi_i\} :  \mathrm{Tr}[\hat\rho \, \hat a^\dagger \hat a]\le \nu, \mathrm{~with~} \hat \rho= \displaystyle\sumint_i p_i \hat \psi_i  \right\}
\end{equation}

Note that, as a consequence of the concavity of the von~Neumann entropy, it is sufficient to consider ensembles of pures states $\hat\psi_i$ in this maximization. 
Further, we may actually restrict the maximization to ensembles $\{p_i,\hat \psi_i\}$ such that $\hat \rho$ has a zero mean vector and a covariance matrix verifying $\mathrm{Tr}[\mathrm{Cov}\left(\hat{\rho}\right)]=1+2\nu$, that is, whose average photon number reaches $\nu$.


\subsection{Capacity of the Gaussian-equivalent channel}

We will compare ${\cal M}$ with the Gaussian-equivalent map denoted as ${\cal M}_G$, in which the environment is prepared in the Gaussian state $\hat \sigma_G$ such that $\mathrm{Cov}\left(\hat{\sigma}_G\right)=\mathrm{Cov}\left(\hat{\sigma}\right)$, namely 
\begin{equation}
{\cal M}_G:\quad \hat\rho \to {\cal M}_G[\hat\rho] = \hat\rho \, \boxplus_\eta  \, \hat\sigma_G
\coloneqq \Tr_2[\hat{U}_\eta(\hat\rho\otimes\hat\sigma_G)\hat{U}^\dagger_\eta]
\end{equation}
Since we restrict to $\mathrm{Cov}\left(\hat{\sigma}\right)\propto\mathbf{I}_2$, the environment state is such that $\mathrm{Cov}\left(\hat{\sigma}_G\right)\propto\mathbf{I}_2$, so it is a thermal state with the same average photon number as $\hat\sigma$ (which we take equal to $\bar n$). Hence, the resulting Gaussian map ${\cal M}_G$ is phase-insensitive, and it capacity is known \cite{Giovannetti2014}. Since
\begin{equation}
\mathrm{Cov}\left({\cal M}_G[\hat{\rho}]\right) =  
\eta \, \mathrm{Cov}\left(\hat{\rho}\right) + (1-\eta) \, \mathrm{Cov}\left(\hat{\sigma}_G\right)
\end{equation}
it is important to stress that injecting the same state $\hat \rho$ into channels ${\cal M}$ or ${\cal M}_G$ results into two distinct output states that have nevertheless the same covariance matrix $\mathrm{Cov}\left({\cal M}_G[\hat{\rho}]\right) =\mathrm{Cov}\left({\cal M}[\hat{\rho}]\right)$.


When computing the  capacity of ${\cal M}_G$, we know that the maximum in Eq. \eqref{eq:maximum_holevo_information} is attained when the input ensemble is a thermal state $\hat{\tau}_\nu$ of average photon number $\nu$ realized with Gaussian-distributed coherent states $\hat\varphi_\alpha=\proj{\alpha}$, which we denote as $\{p_\alpha,\hat \varphi_\alpha\}$ \cite{Giovannetti2014}. Thus, 
\begin{eqnarray}
C ({\cal M}_G)&=& \bigchi \left[\{p_\alpha,\hat \varphi_\alpha\}, {\cal M}_G \right] \nonumber \\ 
&=& S\left({\cal M}_G[\sumint_\alpha p_\alpha \hat \varphi_\alpha ]\right)-\sumint_\alpha p_\alpha \, S\left({\cal M}_G\left[\hat \varphi_\alpha \right]\right)
\nonumber \\
&=& g(\eta\nu+ \bar n) - g(\bar n)
\end{eqnarray}
where $g(x)=(x+1)\ln(x+1)-x\ln(x)$ is the von Neumann entropy of a thermal state of average photon number $x$. We will now exploit the knowledge of $C ({\cal M}_G)$ in order to lower and upper bound $C ({\cal M})$,
similarly as in the classical case.

Let us note first that the quantum relative entropy between  $\hat\rho$ and $\hat\rho_G$ can be expressed as
\begin{equation}
 D(\hat\rho||\hat\rho_G)=S(\hat\rho_G)-S(\hat\rho) \ge 0
\end{equation}
as long as $\hat\rho_G$ is the Gaussian state with the same covariance matrix and same mean vector as $\hat\rho$ (here we set the mean vector of all states to zero). Just like with the classical relative entropy, this inequality expresses that the Gaussian state $\hat\rho_G$ has the highest von Neumann entropy among all states $\hat\rho$ with the same covariance matrix. Hence, $D(\hat\rho||\hat\rho_G)$ can be viewed as a measure of the non-Gaussianity of $\hat\rho$.

\subsection{Lower bound on $C ({\cal M})$}
We know that the input ensemble $\{p_\alpha,\hat \varphi_\alpha\}$ realizing a thermal state $\hat{\tau}_\nu$ of average photon number $\nu$  has no reason to achieve the capacity $C ({\cal M})$ but it yields  a lower bound to this capacity, namely
\begin{eqnarray}
C ({\cal M}) \ge \bigchi \left[\{p_\alpha,\hat \varphi_\alpha\}, {\cal M} \right] .
\label{eq:trivial_lowerbound_coherent_capacity}
\end{eqnarray}
As for classical channels, we compare the situations where the same input ensemble $\{p_\alpha,\hat \varphi_\alpha\}$ is injected either in channel ${\cal M}$ or in the Gaussian-equivalent channel ${\cal M}_G$. We have
\begin{eqnarray}
\label{eq:chain-ineq-lower-bound-quantum}
\lefteqn {
\bigchi \left[\{p_\alpha,\hat \varphi_\alpha\}, {\cal M} \right] - \bigchi \left[\{p_\alpha,\hat \varphi_\alpha\}, {\cal M}_G \right] 
 } \hspace{1cm} \nonumber \\
&=& S\left({\cal M}[\sumint_\alpha p_\alpha \hat \varphi_\alpha]\right)-\sumint_\alpha p_\alpha S\left({\cal M}\left[\hat \varphi_\alpha \right]\right)
-S\left({\cal M}_G[\sumint_\alpha p_\alpha \hat \varphi_\alpha]\right)+\sumint_\alpha p_\alpha S\left({\cal M}_G\left[\hat \varphi_\alpha \right]\right)
\nonumber \\
&=& \sumint_\alpha p_\alpha S\left({\cal M}_G\left[\hat \varphi_\alpha \right]\right)-\sumint_\alpha p_\alpha S\left({\cal M}\left[\hat \varphi_\alpha \right]\right) -S\left({\cal M}_G[\sumint_\alpha p_\alpha \hat \varphi_\alpha]\right) + S\left({\cal M}[\sumint_\alpha p_\alpha \hat \varphi_\alpha ]\right)
\nonumber \\ 
&=& \sumint_\alpha p_\alpha \, D\Big({\cal M}\left[\hat \varphi_\alpha \right] ~||~ {\cal M}_G\left[\hat \varphi_\alpha \right] \Big)
- D \left( {\cal M}[\sumint_\alpha p_\alpha \hat \varphi_\alpha ] ~||~ {\cal M}_G[\sumint_\alpha p_\alpha \hat \varphi_\alpha ] \right)
\nonumber \\ 
&\ge& 0 
\end{eqnarray}
where we can write quantum relative entropies since applying the maps ${\cal M}$ and ${\cal M}_G$ to a same input state gives two output states with the same covariance matrix.  The inequality in \eqref{eq:chain-ineq-lower-bound-quantum} comes from the double convexity of the quantum relative entropy, namely
\begin{equation}
 D\Big(\sum\nolimits_i p_i \hat \rho_i|| \sum\nolimits_i p_i \hat \sigma_i\Big) \le \sum\nolimits_i p_i \, D(\hat \rho_i|| \hat \sigma_i)
\end{equation}
Thus, combining Eqs. \eqref{eq:trivial_lowerbound_coherent_capacity} and \eqref{eq:chain-ineq-lower-bound-quantum}, we have
\begin{eqnarray}
C ({\cal M}) \ge \bigchi \left[\{p_\alpha,\hat \varphi_\alpha\}, {\cal M} \right] \ge  C ({\cal M}_G)
\end{eqnarray}
where we have used $C ({\cal M}_G) = \bigchi \left[\{p_\alpha,\hat \varphi_\alpha\}, {\cal M}_G \right]$ since the capacity of ${\cal M}_G$ is attained by using the Gaussian encoding $\{p_\alpha,\hat \varphi_\alpha\}$. 
Similarly as for classical channels, the capacity of the Gaussian-equivalent channel $C ({\cal M}_G)$ is thus a lower bound on $C ({\cal M})$. This lower bound is achievable since injecting $\{p_\alpha,\hat \varphi_\alpha\}$ into channel ${\cal M}$ gives the rate $\bigchi \left[\{p_\alpha,\hat \varphi_\alpha\}, {\cal M} \right]$, which cannot be lower than the lower bound $C ({\cal M}_G)$.

\subsection{Upper bound on $C ({\cal M})$}

Just as in the classical analysis, we now compare the situation when some arbitrary ensemble $\{p_i,\hat \psi_i\}$ is injected in the channel ${\cal M}$ and when the Gaussian ensemble $\{p_\alpha,\hat \varphi_\alpha\}$ with the same total covariance matrix is injected in the Gaussian-equivalent channel ${\cal M}_G$:
\begin{eqnarray}
\label{eq:chain-ineq-upper-bound-quantum}
\lefteqn {
\bigchi \left[\{p_i,\hat \psi_i\},{\cal M}\right] - \bigchi \left[\{p_\alpha,\hat \varphi_\alpha\},{\cal M}_G \right] 
 } \nonumber \\
&=& S\left({\cal M}[\sumint_i p_i \hat \psi_i]\right) -\sumint_i p_i S\left({\cal M}[\hat \psi_i]\right)
-S\left({\cal M}_G[\sumint_\alpha p_\alpha \hat \varphi_\alpha]\right)+\sumint_\alpha p_\alpha S\left({\cal M}_G\left[\hat \varphi_\alpha \right]\right)
\nonumber \\
&=& \sumint_\alpha p_\alpha S\left({\cal M}_G\left[\hat \varphi_\alpha \right]\right)-\sumint_i p_i S\left({\cal M}[\hat \psi_i]\right)
-S\left({\cal M}_G[\sumint_\alpha p_\alpha \hat \varphi_\alpha]\right)+ S\left({\cal M}[\sumint_i p_i \hat \psi_i]\right)
\nonumber \\
&=& \sumint_\alpha p_\alpha S\left( {\cal M}_G\left[\hat \varphi_\alpha \right]\right)- \sumint_i p_i S\left( {\cal M}[\hat \psi_i]\right) - D\Big( {\cal M} \left[ \hat \rho \right] ~||~  {\cal M}_G\left[\hat \rho_G\right] \Big)
\nonumber \\ 
&\le& \sumint_\alpha p_\alpha S\left( {\cal M}_G\left[\hat \varphi_\alpha \right]\right)- \sumint_i p_i S\left( {\cal M}[\hat \psi_i ]\right) 
\end{eqnarray}
where we defined $\hat\rho= \displaystyle\sumint_i p_i \hat \psi_i $ and $\hat\rho_G = \displaystyle\sumint_\alpha p_\alpha \hat \varphi_\alpha $, the latter state being the Gaussian state with the same covariance matrix as $\hat\rho$, that is, $\mathrm{Cov}\left(  \hat\rho_G \right)=\mathrm{Cov}\left(  \hat\rho \right)$. The inequality in Eq. \eqref{eq:chain-ineq-upper-bound-quantum} comes from $D( {\cal M} \left[ \hat\rho \right] ~||~  {\cal M}_G\left[\hat\rho_G\right] )\ge 0$, where we can indeed write a relative entropy because $\mathrm{Cov}\left( {\cal M} \left[ \hat\rho \right] \right)= \mathrm{Cov}\left( {\cal M}_G \left[ \hat\rho_G \right] \right)$. Thus, defining 
\begin{equation}
 \Delta [ \{p_i,\hat \psi_i\} ] \coloneqq \sumint_\alpha p_\alpha S\left( {\cal M}_G\left[\hat \varphi_\alpha \right]\right)- \sumint_i p_i S\left( {\cal M}[\hat \psi_i ]\right) ,
 \label{eq:def-delta}
\end{equation}
we have
\begin{equation}
\bigchi \left[\{p_i,\hat \psi_i\},{\cal M}\right] \le \bigchi \left[\{p_\alpha,\hat \varphi_\alpha\},{\cal M}_G \right]
+  \Delta [ \{p_i,\hat \psi_i\} ] ,
\label{eq:precurs-quant-upp-bound}
\end{equation}
which is the counterpart of Eq. \eqref{eq:precurs-class-upp-bound}. To make this analogy more obvious, one should write explicitly the classical analog of Eq. 
\eqref{eq:def-delta}, namely
\begin{eqnarray}
\Delta_{\text{cl}} &=& \!\! \int\!\!\! dx \,\, p_{X_G}(x) \, h(x+N_G|x) - \! \int\!\!\! dx \,\, p_X(x) \, h(x+N|x)  \nonumber \\
&=& h(X_G+N_G|X_G) - h(X+N|X) \nonumber \\
&=& h(N_G) - h(N) \nonumber \\
&=& D(N||N_G)
\label{eq:classical-analogy}
\end{eqnarray}
Nicely enough, $\Delta_{\text{cl}}$ does not depend on the input probability density $p_X$ for classical channels.
This results from the fact that the noise is additive and the Shannon differential entropy is invariant under translation, that is, $h(X+a)=h(X)$, $\forall a\in \mathbb{R}$. This is unfortunatey not so simple for quantum channels \cite{Guha2016-xe}.

In order to get the capacity $C ({\cal M})$, we must maximize 
$\bigchi \left[\{p_i,\hat \psi_i\},{\cal M} \right]$ over all input ensembles $\{p_i,\hat \psi_i\} \in {\cal E}_\nu$. Note first that the Gaussian ensemble $\{p_\alpha,\hat \varphi_\alpha\}$ appearing in Eqs. \eqref{eq:def-delta}-\eqref{eq:precurs-quant-upp-bound} indirectly depends on the considered  
input ensemble $\{p_i,\hat \psi_i\}$ since both ensembles must give an average state (resp. $\rho_G$ and $\rho$) that has the same covariance matrix. This is why we note $\Delta$ as a functional of $\{p_i,\hat \psi_i\}$ only. More precisely, when maximizing Eq.~\eqref{eq:precurs-quant-upp-bound} over $\{p_i,\hat \psi_i\} \in {\cal E}_\nu$, the corresponding maximization over 
the Gaussian-equivalent ensembles $\{p_\alpha,\hat \varphi_\alpha\}$ reaches its maximum  for $\mathrm{Cov}\left(\hat{\rho}_G\right)\propto \mathbf{I}_2$, which only depends on the average photon number $\nu$ since $\mathrm{Tr}[\mathrm{Cov}\left(\hat{\rho}_G\right)]=1+2\nu$.\footnote{Here again, we assume that the input photon number $\nu$ is above some threshold, so that none of the quadratures of $\hat \rho_G$ is squeezed. This is needed since $\hat \rho_G$ is realized with a mixture of coherent states $\{p_\alpha,\hat \varphi_\alpha\}$. We restrict to this regime here.} Thus, the maximization of $\bigchi \left[\{p_\alpha,\hat \varphi_\alpha\}, {\cal M}_G \right]$ in Eq.~\eqref{eq:precurs-quant-upp-bound} yields the capacity of the Gaussian-equivalent channel ${\cal M}_G$, so we get
\begin{equation}
    C ({\cal M}) \le  C ({\cal M}_G) + \max_{\displaystyle \{p_i,\hat \psi_i\} \in {\cal E}_\nu} \Delta [ \{p_i,\hat \psi_i\} ] .
\end{equation}

Importantly, the first term of $\Delta [ \{p_i,\hat \psi_i\} ]$ does not need to be maximized (it is a constant) since $\hat \rho_G$ is known to be realized with coherent states $\hat \varphi_\alpha$ in the maximization of $\bigchi \left[\{p_\alpha,\hat \varphi_\alpha\}, {\cal M}_G \right]$ giving $C ({\cal M}_G)$. Maximizing the second term amounts to finding the minimum output entropy of the non-Gaussian channel ${\cal M}$ (averaged over the input symbols). The latter entropy can in turn be further lower bounded by finding the pure state $\hat\psi$ that minimizes the output entropy $S\left( {\cal M}\left[\hat \psi \right]\right)$ since the von Neumann entropy is concave. Hence, the
maximum of $\Delta [ \{p_i,\hat \psi_i\} ]$ can itself be upper bounded, resulting in our final upper bound on the capacity
\begin{equation}
    C ({\cal M}) \le  C ({\cal M}_G) + \Delta_\mathrm{max}
    \label{eq:quant-upp-bound}
\end{equation}
where the potential capacity increase is defined as
\begin{equation} \Delta_\mathrm{max} =  S_\mathrm{min}^{{\cal M}_G} - S_\mathrm{min}^{{\cal M}}
\label{def:delta-max}
\end{equation}
with
\begin{eqnarray}
&& S_\mathrm{min}^{{\cal M}_G}  = \min_{\displaystyle \hat\psi}  S\left( {\cal M}_G[\hat \psi] \right) \equiv S\left( {\cal M}_G\left[\ket{0}\bra{0} \right]\right) \nonumber \\
&& S_\mathrm{min}^{{\cal M}} = \min_{\displaystyle \hat\psi}  S\left( {\cal M}[\hat \psi] \right)
\end{eqnarray}
Here, $S_\mathrm{min}^{{\cal M}}$ stands for the minimum output entropy of the non-Gaussian channel ${\cal M}$, while $S_\mathrm{min}^{{\cal M}_G}$ stands for the minimum output entropy of the Gaussian-equivalent channel ${\cal M}_G$, which is achieved simply by inputting the vacuum state $\ket{0}$ (or any coherent state $\hat\varphi_\alpha$).

\subsection{Properties of $\Delta_\mathrm{max}$}

The potential capacity increase $\Delta_\mathrm{max}$ is thus expressed as an entropy difference, which plays a similar role as the relative entropy $D(N || N_G)$ for classical non-Gaussian channels. Indeed, a classical input signal can simply be fully zeroed in order to minimize the output entropy since there is no quantum noise, that is, $p_X(x)=\delta(x)$, with $\delta$ denoting the Dirac distribution. Accordingly, we may thus rewrite Eq. \eqref{eq:classical-analogy} as
\begin{eqnarray}
\Delta_{\text{cl}} = D(N || N_G) 
= h({\cal M}_G[\delta]) - h({\cal M}[\delta]) 
\label{eq:classical-analogy-2}
\end{eqnarray}
which is the straightforward analog of Eq.~\eqref{def:delta-max}. In the quantum case, however, we have to express the difference between the (averaged) output entropies when considering the optimal ensemble of input pure states.

As an interesting special case, consider the non-Gaussian channels ${\cal M}$ that are such that the minimum output entropy is reached by the vacuum state $\ket{0}$ (or any coherent state $\hat\varphi_\alpha$), just as for the Gaussian-equivalent channels ${\cal M}_G$. If ${\cal M}$ satisfies this property of having a Gaussian minimizer (such cases will be discussed in Sec.~\ref{sec:conjectures}), then its potential capacity increase can be written as 
\begin{eqnarray}
\Delta_\mathrm{max} 
&=& S\left( {\cal M}_G\left[\ket{0}\bra{0} \right]\right) - S\left( {\cal M}\left[\ket{0}\bra{0} \right]\right) \nonumber \\
&=& D \Big( {\cal M}\left[\ket{0}\bra{0} \right] ~||~ {\cal M}_G\left[\ket{0}\bra{0} \right] \Big)
\end{eqnarray}
which is simply the non-Gaussianity of the output state of the channel ${\cal M}$ when the input is taken as the minimizing state 
$\ket{0}$ (or any coherent state $\hat\varphi_\alpha$). In this case, the expression is identical to its classical counterpart $\Delta_{\text{cl}}$, see Eq.~\eqref{eq:classical-analogy} or
\eqref{eq:classical-analogy-2}.

More generally, if vacuum is not the state minimizing the output entropy of $\mathcal{M}$, $\Delta_{\mathrm{max}}$ is still easily shown to be non-negative.
Indeed, we have
\begin{align}
\Delta_\mathrm{max} 
&=
S\left( {\cal M}_G[\ket{0}\bra{0}] \right)
-
\min_{\displaystyle \hat\psi}  S\left( {\cal M}[\hat \psi] \right)
\nonumber
\\&\geq
S\left( {\cal M}_G[\ket{0}\bra{0}] \right)
-
S\left( {\cal M}[\ket{0}\bra{0}] \right),
\end{align}
which is always non-negative as a consequence of our previous argument.
The non-Gaussianity of the output associated to vacuum can thus always be used as a lower bound to the value of $\Delta_\mathrm{max}$.

Note, finally, that the upper bound \eqref{eq:quant-upp-bound} becomes tight in the limit where $\nu\gg 1$. Indeed, the capacity 
$C ({\cal M})$ tends to the upper bound $C ({\cal M}_G) + \Delta_\mathrm{max}$ in this case since the averaged output state of ${\cal M}$ can be made to approach a thermal (Gaussian) state simply by choosing a Gaussian input ensemble. In contrast, the upper bound \eqref{eq:quant-upp-bound} becomes loose in the limit of $\nu\ll 1$. This is because $\Delta_\mathrm{max}$ is a characteristic of the channel ${\cal M}$ and is independent of $\nu$. When $\nu\to 0$, the upper bound on $C ({\cal M})$ tends to a constant value $\Delta_\mathrm{max}$ as $C ({\cal M}_G)\to 0$, whereas we expect that the capacity 
$C ({\cal M})\to 0$.

\subsection{Quantum amplifier channel}
\label{sec:quantum-amplifier-channel}
The exact same treatment extends to quantum amplifier channels with non-Gaussian noise, corresponding to $\eta>1$. This is because the reasoning is independent of $\eta$ and only depends on the way the input and output covariance matrices are related. The same is true for phase-conjugating channels, corresponding to $\eta<0$. 
Hence, the upper and lower bounds on the capacity hold for all non-Gaussian attenuator and (phase-conjugate) amplifier channels with an arbitrary environment state $\sigma$.

\section{Minimum output entropy: numerical exploration }
\label{sec:conjectures}

The difference between the upper and lower bounds on the capacity 
$C ({\cal M})$ that we derived in Section \ref{Sec:quantum} is a difference between the minimum output entropy of the non-Gaussian channel ${\cal M}$ and its Gaussian associate ${\cal M}_G$.
The minimum output entropy (MOE) of Gaussian channels has been extensively studied in the literature \cite{Giovannetti2004-ns, Giovannetti2004-zl, Giovannetti2014}; however very little is known for non-Gaussian channels \cite{Memarzadeh2016-cc}.

In the present section, we investigate the MOE of quantum attenuator channels with non-Gaussian environment (it is expected that the case of quantum  amplifier channels would lead to the same conclusions).
We first consider the case of an environment that is phase-invariant, and then look at a particular example of an environment that possesses other relevant symmetries. To help us identifying the input states that minimize the output entropy, we make use of a numerical minimization routine described in Appendix \ref{apd:min_routine}. We stress that our findings in this section are thus based on numerics only and do not lie on analytical results.
However, supported by these numerics, we suggest and conjecture some properties for the states minimizing the output entropy of quantum attenuators. In a nutshell, we make the following observations (detailed in the rest of this section):
\begin{itemize}
    \item 
    Coherent states achieve the MOE of quantum attenuators with phase-invariant (\textit{i.e.}, Fock-diagonal) environments.
    \item 
    The MOE state of a quantum attenuator is, in some cases, not unique (even if the trivial non-uniqueness due to displacements is disregarded).
    \item 
    The MOE state of a quantum attenuator is, in some cases, non-Gaussian (this can only happen if the environment is non-Gaussian).
    \item 
    The MOE state of a quantum attenuator is invariant under the phase-space symmetries respected by the environment.
    \item 
    For a fixed environment, the MOE state of a quantum attenuator may depend on the transmittance $\eta$.
\end{itemize}

\subsection{Preliminaries: properties of quantum attenuator channels}
\label{subsec:prop_quantum_attenuator}

\begin{figure}[t]
\includegraphics[width=\linewidth]{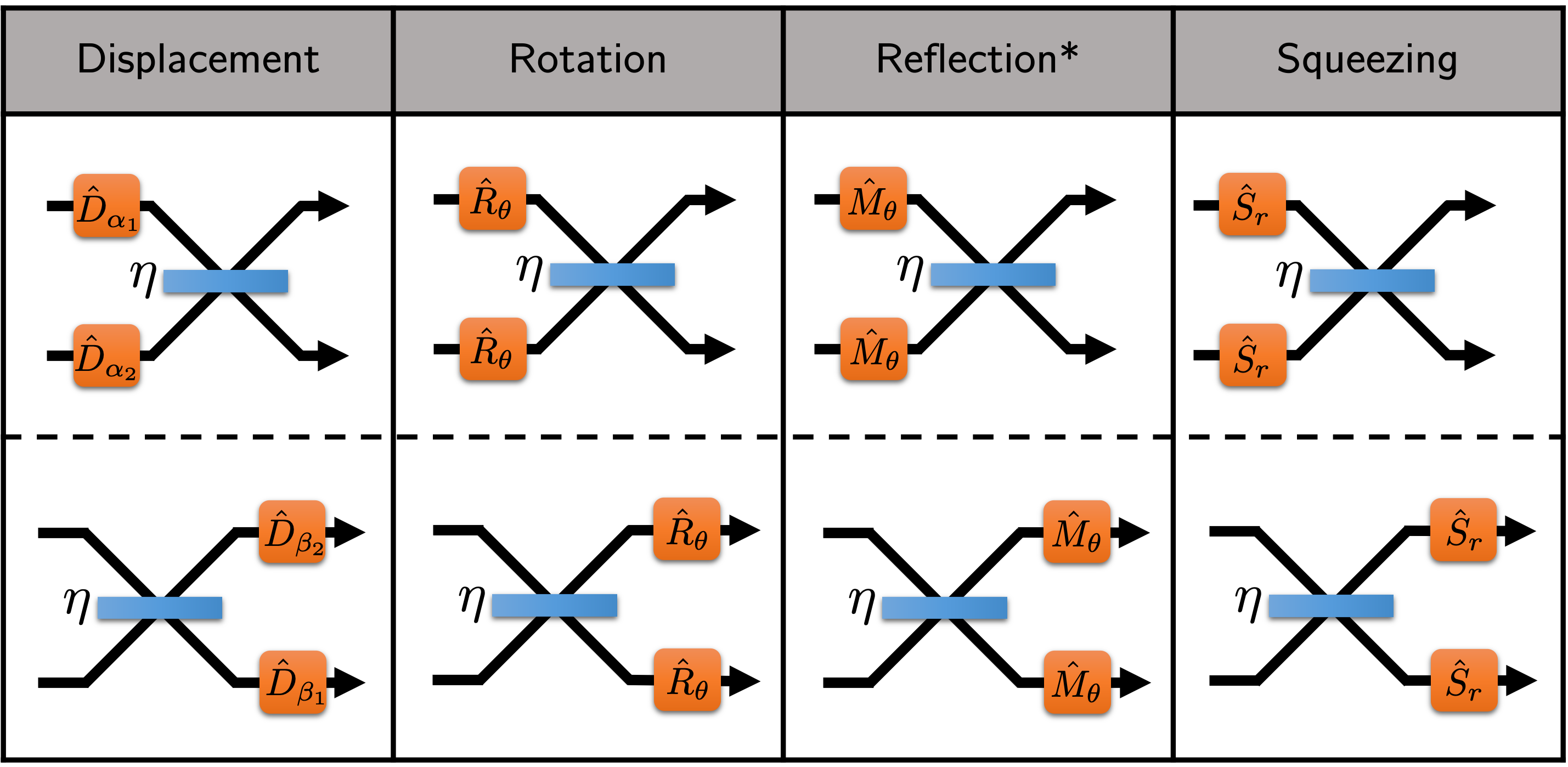}
\caption{
Commutation properties of the beam splitter.
For each column, the upper and lower setups yield the same two-mode unitary operation (up to a global phase).
In the first column (displacement),  the relation holds provided $\beta_1=\sqrt{\eta}\,\alpha_1+\sqrt{1-\eta}\,\alpha_2$ and $\beta_2=-\sqrt{1-\eta}\,\alpha_1+\sqrt{\eta}\,\alpha_2$.
In the second column (rotation), the commutation relation holds provided all rotation operators have the same angle $\theta$.
In the third column (reflection), the commutation relation holds provided all reflection operators have the same angle $\theta$.
$(^\ast$~Note that reflections are not physically implementable since they are related to the phase-conjugation  operator, which is anti-unitary.)
In the fourth column (squeezing), the commutation relation holds provided all squeezing operators have the same squeezing parameter $r$.
}
\label{fig:beamsplitter_commutation}
\end{figure}

Before diving into the results of our numerical exploration, let us recall some important properties of the family of (generalized) quantum attenuators.
The unitary transformation implemented by the beam splitter benefits from several commutation relations with Gaussian unitary operators, which carry on to the corresponding channels.
These commutation rules are illustrated in Fig. \ref{fig:beamsplitter_commutation}.

First,  a displacement operation on the inputs of a beam splitter is equivalent to a displacement on its outputs with the adequate displacement parameters.
A direct consequence of that equivalence is that applying the operator $\hat{D}_\alpha$ to the input of a quantum attenautor with transmittance $\eta$ produces the same result as applying the operator $\hat{D}_{\sqrt{\eta}\,\alpha}$ to its output.
More precisely, we may write
\begin{align}
\mathcal{M}\left[
\hat{D}_\alpha
\,
\hat{\rho}
\,
\hat{D}^\dagger_\alpha
\right]
=
\hat{D}_{\sqrt{\eta}\,\alpha}
\,
\mathcal{M}\big[\hat{\rho}\big]
\,
\hat{D}_{\sqrt{\eta}\,\alpha}^\dagger,
\end{align}
for any quantum attenuator $\mathcal{M}$ with transmittance $\eta$ (and arbitrary environment).
Since displacement is a unitary operator, this also means that the eigenvalues of the output are preserved under any displacement of the input, hence the output always possesses the same von Neumann entropy.
When looking for the state that minimize the output entropy of a quantum attenuator, it is thus sufficent to restrict to inputs with zero mean displacement.
For the same reason, we may, without loss of generality, only consider environments with zero mean displacement.

Second, rotating both inputs of a beam splitter is equivalent to rotating its two outputs as long as the rotation angles are the same.
This property is illustrated in the second column of Fig. \ref{fig:beamsplitter_commutation}.
A similar commutation relation holds for the reflection operator $\hat{M}_\theta$ that we define as follows.
Denote as $\hat{M}$ the phase-conjugation operation (the letter $M$ is chosen as $\hat{M}$ acts as a \textit{mirror} over the $x$ axis in phase space).
The map $\ket{\psi}\rightarrow\hat{M}\ket{\psi}$ corresponds to $\psi(x)\rightarrow\psi^\ast(x)$ and $W(x,p)\rightarrow W(x,-p)$.
We then define the reflection operator with angle $\theta$ as $\hat{M}_\theta=\hat{R}_\theta\hat{M}\hat{R}^\dagger_\theta$.
In phase space, the operator $\hat{M}_{\theta}$ performs a reflection around an axis intersecting the origin of phase-space and making an angle $\theta$ with the $x$-axis.
The commutation between the reflection operator and beam splitter is illustrated in the third column of Fig.~\ref{fig:beamsplitter_commutation}.

With the above observations in mind, we introduce the following lemma:
\begin{lem}[Channel covariance]
A quantum attenuator channel is covariant with respect to the  rotation and reflection symmetries of its environment.
Let $\mathcal{M}$ be a quantum attenuator with transmittance $\eta$ and environment $\hat{\sigma}$.
We then have:
\begin{align*}
\forall\hat{R}_\theta\;|\;
\hat{\sigma}=
\hat{R}_\theta\hat{\sigma}\hat{R}^\dagger_\theta
\ \,:&\ \,
\quad
\mathcal{M}[\hat{R}_\theta\hat{\rho}\hat{R}_\theta^\dagger]
=
\hat{R}_\theta
\mathcal{M}[\hat{\rho}]
\hat{R}_\theta^\dagger,
\\
\forall\hat{M}_\theta\;|\;
\hat{\sigma}=
\hat{M}_\theta\hat{\sigma}\hat{M}^\dagger_\theta:
&
\quad
\mathcal{M}[\hat{M}_\theta\hat{\rho}\hat{M}_\theta^\dagger]
=
\hat{M}_\theta
\mathcal{M}[\hat{\rho}]
\hat{M}_\theta^\dagger.
\end{align*}
\label{lem:commutation_quantum_attenuator}
\end{lem}
Lemma \ref{lem:commutation_quantum_attenuator} is a direct consequence of the commutation relations between the beam splitter unitary and the rotation or reflection operators as schematized in Fig.~\ref{fig:beamsplitter_commutation}.
We have also used the fact that partial tracing is invariant under a unitary operation over the partial-traced mode.

Let us finally highlight a last commutation rule between the beam splitter and the squeezing operator.
Squeezing both inputs of the beam splitter is equivalent to squeezing its two outputs as long as the squeezing parameters are the same.
This relation is illustrated in the fourth column of Fig. \ref{fig:beamsplitter_commutation}.
Combining this relation with the commutation rule between the beam-splitter unitary and the rotation operator, it appears that we may restrict, when looking for minimum-output-entropy states, to environments with a covariance matrix proportional to the identity.
If the environment has a covariance matrix in a different form, it is always possible to use the appropriate rotation and squeezing operators in order to make its covariance matrix proportional to the identity matrix.

In conclusion, we may consider, without loss of generality, environments which have a zero mean displacement and a covariance matrix proportional to the identity matrix.
The environment $\hat{\sigma}$ has a zero mean displacement if and only if $\Tr[\hat{\sigma}\hat{a}]=0$.
A simple derivation gives $\hat{x}\hat{p}+\hat{p}\hat{x}=i\big(\hat{a}^{\dagger 2}-\hat{a}^2\big)$ and $\hat{x}^2-\hat{p}^2=\hat{a}^{\dagger 2}+\hat{a}^2$.
As a consequence, any zero-mean environment $\hat{\sigma}$ has a covariance matrix proportional to the identity if and only if $\mathrm{Tr}[\hat{\sigma}\hat{a}^2]=0$ .
Note that we will also only consider input states $\hat{\rho}$ such that $\mathrm{Tr}[\hat{\rho}\hat{a}]=0$, but 
the input states may have an arbitrary covariance matrix, so that $\mathrm{Tr}[\hat{\rho}\hat{a}^2]$ will in general be nonzero.


\subsection{Phase-invariant environment}

Here, we consider quantum attenuator channels that are associated with a phase-invariant environment.
Recall that a quantum state $\hat{\sigma}$ is said to be phase-invariant if it is invariant under any rotation operation, \textit{i.e.}, such that $\hat{R}_\theta\hat{\sigma}\hat{R}^\dagger_\theta=\hat{\sigma},\ \forall\theta$.
As a consequence of Lemma \ref{lem:commutation_quantum_attenuator}, a quantum attenuator with phase-invariant environment is covariant with respect to any rotation.
Such a channel $\mathcal{M}$ is called phase-covariant (or rotation-covariant), \textit{i.e.} $\mathcal{M}[\hat{R}_\theta\hat{\rho}\hat{R}^\dagger_\theta]=\hat{R}_\theta\mathcal{M}[\hat{\rho}]\hat{R}^\dagger_\theta,\ \forall\theta$.

Let us first focus on phase-covariant quantum attenuators that are associated with a pure environment.
Such channels are thus quantum attenuators with  the environment in a Fock state, and in the following we refer to them as Fock attenuators (also called photon-added channels in Ref. \cite{PhysRevA.95.062309}).
A Fock attenuator with transmittance $\eta$ and the environment in the Fock state $|n\rangle$ is defined as 
\begin{align}
    \mathcal{M}_{\eta,n}
    \big[\hat{\rho}\big]
    =\Tr_2
    \Big[\hat{U}_\eta
    \big(\hat{\rho}
    \otimes
    \ket{n}\bra{n}
    \big)
    \hat{U}^\dagger_\eta
    \Big].
\end{align}

From our (rather exhaustive)  numerical simulations, it appears that the states that minimize the output entropy of Fock attenuators $\mathcal{M}_{\eta,n}$ are coherent states.
Indeed, when running our minimization routine, we observe that the result always converges to a coherent state.
We observe this behaviour regardless of the  value of $\eta$ and $n$.
With this in mind, we lay the following conjecture:
\begin{conj}[Fock attenuator]
The minimum output entropy of a Fock attenuator channel $\mathcal{M}_{\eta,n}$ is achieved by coherent states, $\forall \eta, n$.
\label{conj:moe_fock_attenuator}
\end{conj}

Some extra observations should be made about Conjecture \ref{conj:moe_fock_attenuator}.
First, when the environment is in the vacuum state, the conjecture is trivially true as the output associated to vacuum $\mathcal{M}_{\eta,0}[\ket{0}\bra{0}]=\ket{0}\bra{0}$ is a pure vacuum state (this is expected as the channel is then a Gaussian channel).

Second, we notice that there sometimes exist other states performing as well as coherent states.
For example, let us consider the Fock attenuator with transmittance $\eta=1/2$ and environment in the Fock state $|1\rangle$, so that $\hat{\sigma}=\ket{1}\bra{1}$.
For that particular channel, the output associated to vacuum is $\mathcal{M}_{1/2,1}(\ket{0}\ket{0})=(\ket{0}\bra{0}+\ket{1}\bra{1})/2$ and the output associated to the Fock state 1 is $\mathcal{M}_{1/2,1}(\ket{1}\bra{1})=(\ket{0}\bra{0}+\ket{2}\bra{2})/2$ (as a consequence of the Hong-Ou-Mandel effect).
These two outputs state have identical eigenvalues $\lbrace 1/2, 1/2\rbrace$ so that they also have equal von Neumann entropies.

\begin{figure}[t]
\centering
\includegraphics[width=0.7\linewidth]{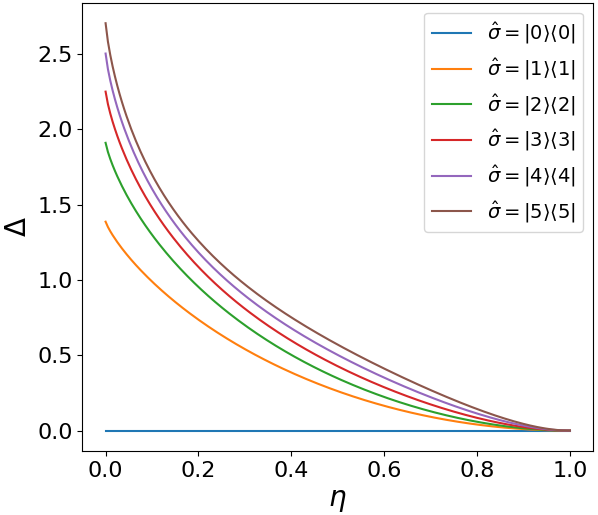}
\caption{
Capacity interval width $\Delta$ for several Fock attenuator channels (with the environment in state $\hat \sigma=\proj{n}$) as a function of the transmittance $\eta$.
We note that  $\Delta$ is independent of the constraint on the photon number $\nu$ at the input.
}
\label{fig:delta_fock_attenuators}
\end{figure}

\begin{figure}[t]
\centering
\includegraphics[width=0.6\linewidth]{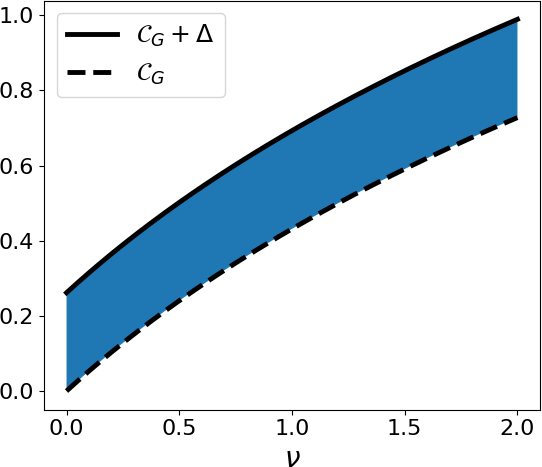}
\caption{
Upper and lower bounds on the classical capacity $\mathcal{C}$ of a non-Gaussian quantum channel as a function of the photon number $\nu$ at the input.
The illustrated example is a Fock attenuator with environment $\hat{\sigma}=\ket{1}\bra{1}$ and transmittance $\eta=1/2$.
The classical capacity $\mathcal{C}$ must lie in the blue area, between the lower bound $\mathcal{C}_G$ and the upper bound $\mathcal{C}_G+\Delta$.
In the regime of low $\nu$, the capacity $\mathcal{C}$ goes to zero so that $\mathcal{C}\rightarrow\mathcal{C}_G$ for $\nu\ll 1$.
In the regime of high $\nu$, the capacity $\mathcal{C}$ tends to the upper bound $\mathcal{C}\rightarrow\mathcal{C}_G+\Delta$ for $\nu\ll 1$ as the bound becomes asymptotically achievable.
}
\label{fig:capacity_interval}
\end{figure}

Assuming Conjecture \ref{conj:moe_fock_attenuator} is true, we are in position to compute the capacity interval width $\Delta$ for Fock attenuators.
Recall that $\Delta=S^{\mathcal{M}_G}_\mathrm{min}-S^{\mathcal{M}}_\mathrm{min}$, so that we can now simply set $S^{\mathcal{M}}_\mathrm{min}$ to the output entropy associated to vacuum $S^\mathcal{M}_0\vcentcolon=S(\mathcal{M}(\ket{0}\bra{0}))$, hence we would have $\Delta = S^{\mathcal{M}_G}_\mathrm{min}-S^{\mathcal{M}}_0$.
Note that, in case Conjecture \ref{conj:moe_fock_attenuator} was proven wrong (so that $S^\mathcal{M}_0>S^{\mathcal{M}}_\mathrm{min}$), the value $S^{\mathcal{M}_G}_\mathrm{min}-S^{\mathcal{M}}_0$ would then be a lower bound on the true value of $\Delta$.
Under the assumption of Conjecture \ref{conj:moe_fock_attenuator}, we illustrate the value of $\Delta$ for Fock attenuators in Fig. \ref{fig:delta_fock_attenuators}.
In Fig. \ref{fig:capacity_interval}, we focus on the particular example of the Fock attenuator with transmittance $\eta=1/2$ and environment $\hat{\sigma}=\ket{1}\bra{1}$ and plot the interval of admissible values of its classical capacity as a function of the photon number constraint at its input.

Let us now extend the discussion to the case of phase-covariant attenuator channels with mixed environments.
Such channels have an environment state $\hat{\sigma}$ that is diagonal in the Fock basis, \textit{i.e.}, such that $\hat{\sigma}=\sum p_n\ket{n}\bra{n}$ for some probability vector $\mathbf{p}$.
A phase-covariant attenuator channel is thus defined as 
\bigskip\begin{align}
    \mathcal{M}_{\eta,\mathbf{p}}
    \big[\hat{\rho}\big]
    =\Tr_2
    \left[\hat{U}_\eta
    \left(\hat{\rho}
    \otimes
    \left(
    \sum\limits_{n=0}^{\infty}
    p_n
    \ket{n}\bra{n}
    \right)
    \right)
    \hat{U}^\dagger_\eta
    \right]
    =
    \sum\limits_{n=0}^{\infty} p_n\ 
    \mathcal{M}_{\eta, n}
    \big[
    \hat{\rho}
    \big],
    \nonumber
\end{align}
so that it is simply a convex mixture of Fock attenuators.

As we did for Fock attenuators, we ran our minimization routine to identify the input states achieving the lowest output entropy for such channels.
For various choices of $\eta$ and $\mathbf{p}$, the routine always converged towards a coherent state.
This leads us to lay the following extension to Conjecture \ref{conj:moe_fock_attenuator}:

\begin{conj}[Phase-covariant attenuator]
 The minimum output entropy of a phase-covariant attenuator channel $\mathcal{M}_{\eta,\mathbf{p}}$ is achieved by coherent states, $\forall \eta, \mathbf{p}$.
\label{conj:moe_covariant_attenuator}
\end{conj}

Conjecture \ref{conj:moe_covariant_attenuator} is a generalization of Conjecture \ref{conj:moe_fock_attenuator} to mixed environments.
Note that in the special case of a thermal environment (\textit{i.e.}, when $\mathbf{p}$ is a geometric distribution), Conjecture \ref{conj:moe_covariant_attenuator} is of course known to hold from Ref. \cite{Giovannetti2014}.

\subsection{Environment with discrete symmetries}

Let us now consider a family of quantum attenuators where,
rather than being phase-invariant, the environment possesses a finite number of rotation and reflection symmetries.
That is, we consider an environment $\hat{\sigma}$ such that there exist a finite set $\lbrace\theta_1,...,\theta_M\rbrace$ and $\lbrace\varphi_1,...,\varphi_N\rbrace$ such that $\hat{R}_{\theta_i}\hat{\sigma}\hat{R}^\dagger_{\theta_i}=\hat{\sigma}$ and $\hat{M}_{\varphi_i}\hat{\sigma}\hat{M}^\dagger_{\varphi_i}=\hat{\sigma}$.
From Lemma \ref{lem:commutation_quantum_attenuator}, this implies that the corresponding quantum attenuator is covariant with respect to these sets of rotations and reflections, but not with respect to any rotation or reflection (hence, it is in general not phase-covariant).

\begin{figure}
\centering
\includegraphics[width=0.7\linewidth]{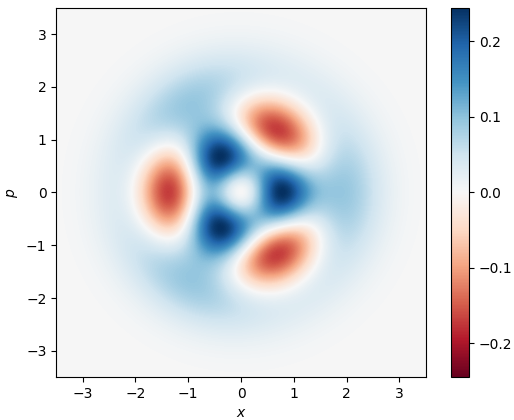}
\caption{
Wigner function of the state $\ket{\psi}=(\ket{0}+\ket{3})/\sqrt{2}$.
The state $\ket{\psi}$ has a zero mean displacement vector and a covariance matrix proportional to the identity.
It has 2 rotation symmetries, namely $\hat{R}_\theta$ for $\theta\in\lbrace{2\pi/3,4\pi/3\rbrace}$, and 3 reflection symmetries, namely $\hat{M}_\theta$ for $\theta\in\lbrace 0,2\pi/3,4\pi/3\rbrace$. We are considering a quantum  attenuator channel whose environment is in state $\ket{\psi}$.
}
\label{fig:wigner_0+3}
\end{figure}

We will only focus here on one particular example of an environment with such discrete symmetries (we expect our conclusions to be common to all channels with  an environment obeying this type of symmetries).
Let us consider the state
\begin{align}
    \ket{\psi}
    =
    \frac{1}{\sqrt{2}}
    \big(
    \ket{0}+\ket{3}
    \big),
    \label{eq:psi_0+3}
\end{align}
 which has a zero mean displacement vector and a covariance matrix proportional to the identity.
The state $\ket{\psi}$ admits a 3-fold rotational symmetry so that it is invariant under $\hat{R}_{\theta}$ for $\theta\in\lbrace2\pi/3, 4\pi/3\rbrace$.
Moreover, since it has real-valued Fock coefficients, it is invariant under phase conjugation.
More precisely, $\ket{\psi}$ is invariant under $\hat{M}_{\theta}$ for $\theta\in\lbrace 0,2\pi/3, 4\pi/3\rbrace$.
Its Wigner function is represented in Fig. \ref{fig:wigner_0+3}.

\begin{figure}
\includegraphics[width=0.8 \linewidth]{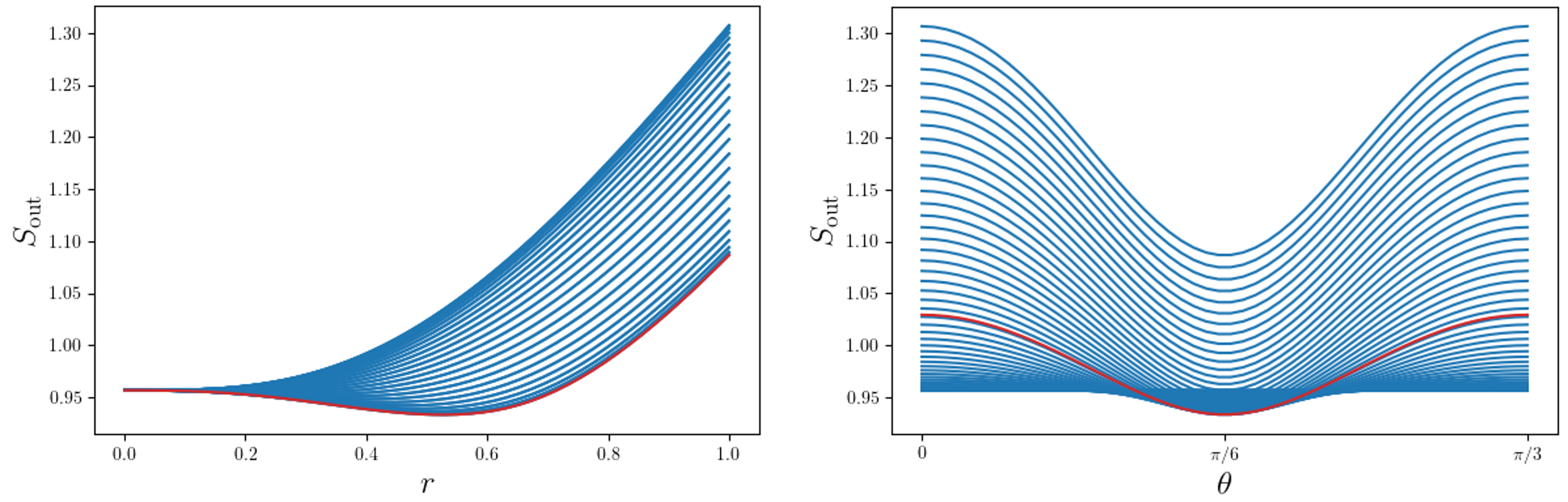}
\centering
\caption{
Output entropy $S_\mathrm{out}$ of the quantum attenuator with transmittance $\eta=1/2$ and environment state $\hat{\sigma}=\ket{\psi}\bra{\psi}$, where $\ket{\psi}=\big(\ket{0}+\ket{3}\big)/\sqrt{2}$.
We compare the output entropy associated to Gaussian squeezed states of the form $\hat{R}_\theta\hat{S}_r\ket{0}$ at the input.
The left plot shows $S_\mathrm{out}$ as a function of $r$, the different blue curves corresponding to different values of $\theta$ (sampled over $[0,\pi/3]$).
The red curve corresponds to the value $\theta=\pi/6$.
Note that the output entropy increases monotonically with $r$ for $r\geq 1$.
The right plot shows $S_\mathrm{out}$  as a function of $\theta$, the different blue curves corresponding to different values of $r$ (sampled over $[0,1]$).
The red curve correspond to the value $r=0.515187$.
It appears that the minimum output entropy (among squeezed states) is achieved for $\theta=\pi/6$ and $r\approx 0.515187$; for that state we find $S_{\mathrm{out}}\approx 0.93333$.
}
\label{fig:output_entropy_squeezed_states}
\end{figure}

Let us now consider the quantum attenuator $\mathcal{M}$ with transmittance $\eta=1/2$ and environment $\hat{\sigma}=\ket{\psi}\bra{\psi}$.
In the previous subsection, we had observed that coherent states appeared to be the states of minimum output entropy for phase-covariant channels.
For the present channel with restricted symmetries, it is thus natural to question whether coherent states (or, more generally, some Gaussian pure states) may again achieve the minimum output entropy.
Remember that to sample the output entropy associated to every Gaussian pure state, it suffices to look at squeezed states with zero displacement, \textit{i.e.}, pure states of the form $\hat{R}_{\varphi}\hat{S}_r\ket{0}$.
Moreover, taken into account the symmetry the environment, it suffices to sample $\varphi$ over $[0, \pi/3]$.
The result of our numerical investigations are plotted in Fig. \ref{fig:output_entropy_squeezed_states}.
It appears that the Gaussian squeezed state with angle $\theta=\pi/3$ and squeezing parameter $r\approx 0.515$ is the state that achieves the minimum output entropy among all Gaussian input states, and the output entropy yields $S_{\mathrm{out}}\approx 0.933$.
It is interesting to notice that the state minimizing the output entropy among Gaussian states is a squeezed state even though the covariance matrix of $\ket{\psi}$ is proportional to the identity.

\begin{figure}[t]
\includegraphics[width=\linewidth]{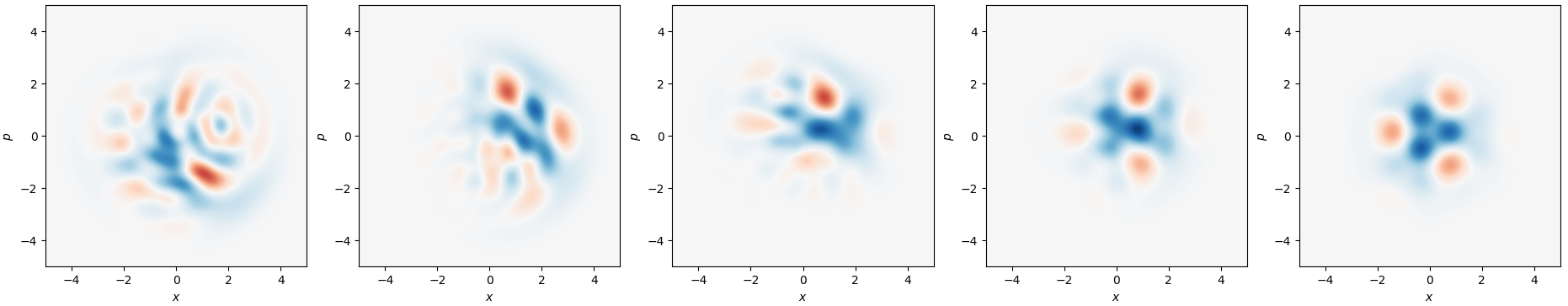}
\caption{
Illustration of the search for the MOE states performed by our minimization routine described in Appendix \ref{apd:min_routine}.
We consider a channel $\mathcal{M}$ with transmittance $\eta=1/2$ and environment state $\hat{\sigma}=\ket{\psi}\bra{\psi}$ for $\ket{\psi}=(\ket{0}+\ket{3})/\sqrt{2}$ (see its Wigner function in Fig. \ref{fig:wigner_0+3}).
Left state is the initial random state.
Then, each step further right correspond to a state of lower output entropy than the previous one, as selected by our routine.
Right state is the final state.
Observe that the last state exhibits the same rotation and reflection symmetries as $\ket{\psi}$.
}
\label{fig:wigner_moe_routine}
\end{figure}

\begin{figure}
\includegraphics[width=\linewidth]{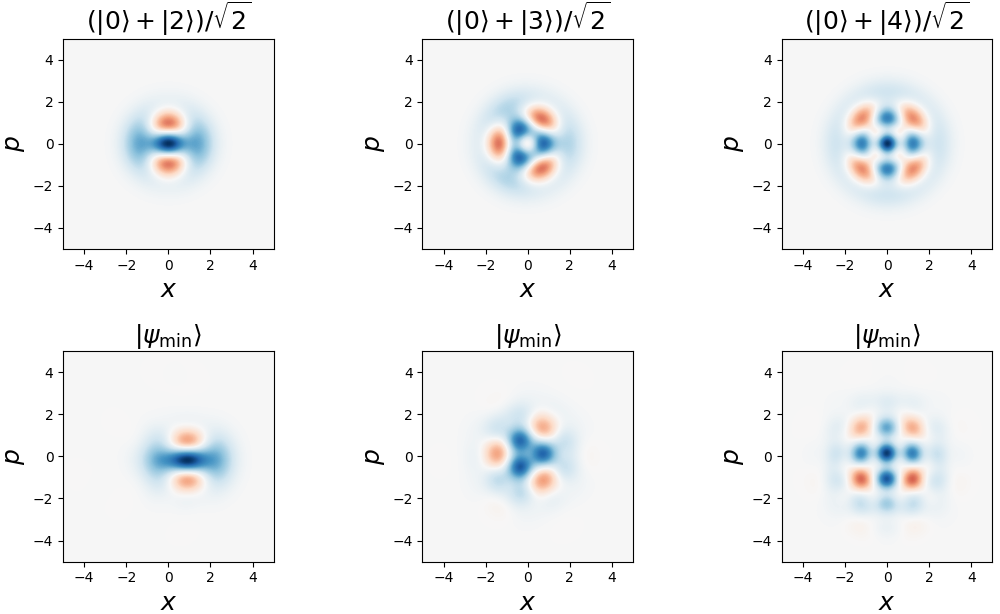}
\caption{
For various environment states (shown on top of each column), we run our routine to find the state that minimizes the output entropy.
The upper row corresponds to the Wigner function of the environment state, and the lower row corresponds to the Wigner function of the input state minimizing the output entropy, as found by our routine. We note that the (discrete) rotation and reflection symmetries are the same.
In each case, the transmittance has been chosen to $\eta=1/2$.
}
\label{fig:moe_various_env}
\end{figure}

We now run our minimization routine and check whether there exists an input state that outperforms the squeezed state identified in Fig. \ref{fig:output_entropy_squeezed_states}. The answer is positive as the routine converges towards a non-Gaussian state that achieves an output entropy $S_\mathrm{out}\approx 0.872$, which is striclty lower than what achieves the optimal Gaussian pure state.
We illustrate our numerical result in Fig. \ref{fig:wigner_moe_routine}.
It is remarkable that the routine converges towards a state 
that shares the same rotation and reflection symmetries as the environment state $\ket{\psi}$.
We have observed a similar behavior for attenuator channels whose environment exhibits other   discrete symmetries and for different values of $\eta$.
We illustrate these findings in Fig. \ref{fig:moe_various_env}.
This leads us to make the following conjecture:
\begin{conj}[Phase-space symmetries]
    The (zero-mean) state achieving the minimum output entropy of a quantum attenuator channel is invariant under the same rotation and reflection symmetries as the environment.
    \label{conj:symmetries}
\end{conj}
\medskip
Observe that Conjecture \ref{conj:symmetries} is consistent with Conjectures \ref{conj:moe_fock_attenuator} and \ref{conj:moe_covariant_attenuator}, but does not imply them.
For example, for phase-covariant attenuators, Conjecture \ref{conj:symmetries} would imply that the minimum output entropy is achieved by any Fock state (while $\ket{0}$ is the only MOE state among them).
We stress that Conjecture \ref{conj:symmetries} only applies to zero-mean states, as only those states may possess a rotation symmetry.

\begin{figure}[t]
\includegraphics[width=\linewidth]{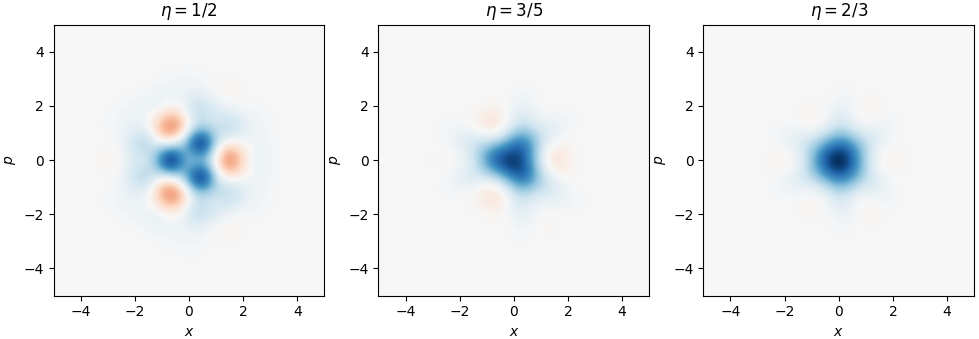}
\caption{
We consider a quantum attenuator with environment $\ket{\psi}=(\ket{0}+\ket{3})/\sqrt{2}$ and various values of transmittance ($\eta\in\lbrace 1/2, 3/5, 2/3\rbrace$).
We numerically minimize the output entropy and observe that the state minimizing the output entropy varies depending on the value of $\eta$. Yet, the MOE state always keeps the exact same rotation and reflection symmetries as the environment state $\ket{\psi}$.
Finally, note that since the channel has a pure environment, it suffices to apply a $\pi$~rotation to the MOE state associated to $\eta$ to obtain the MOE state associated to $1-\eta$.
}
\label{fig:moe_various_eta}
\end{figure}

Let us conclude this analysis by investigating how does the MOE state depend on the transmittance $\eta$. For phase-covariant attenuators, we had observed from numerics that the minimum-output-entropy states are coherent states and that this holds for any phase-invariant environment and any value of $\eta$.
We have now seen that turning to other environment states with restricted symmetries makes the MOE state different from a coherent state.
Let us now analyze what happens if we change the value of the transmittance $\eta$, but keep the same environment.
In Fig.~\ref{fig:moe_various_eta}, we compare three different attenuators channels, each associated with the same environment $\ket{\psi}$ but with different values of transmittance.
 Taking advantage of Conjecture \ref{conj:symmetries} (taken for granted), we can simplify the search for MOE states by restricting to the set of states exhibiting the same symmetries as the environment.
In particular, any state with a $m$-fold rotation symmetry is an eigenstate of the rotation operator $\hat{R}_{2\pi/m}$ and can be written as:
\begin{align}
    \ket{\phi}
    =
    \sum\limits_{n=0}^{\infty}
    c_n\ket{m\cdot n+p}.
    \label{eq:nfold_sym}
\end{align}
where $p\in\mathbb{N}$ is some offset.
We have then $\hat{R}_{2\pi/m}\ket{\phi}=\exp(-2\pi i p/m)\ket{\phi}$.
Adding the constraint that $\ket{\phi}$ has real coefficient in the Fock basis (so that $c_n\in\mathbb{R}\ \forall n$) then ensures that $\ket{\phi}$ also has the corresponding reflection symmetries.
In Fig.~\ref{fig:moe_various_eta}, we present the results of our numerics, where we have used our minimization routine to find the MOE state with an environment state $\ket{\psi}=(\ket{0}+\ket{3})/\sqrt{2}$ and various values of transmittance. Although the MOE state changes, the symmetries survive.

Let us make a final remark concerning Fig. \ref{fig:moe_various_eta}.
Changing the transmittance of a beam-splitter from $\eta$ to $1-\eta$ and adding a phase-rotation of $\pi$ on one of its input is equivalent to permuting the two outputs, up to some rotations at the output.
Moreover, when a quantum attenuator with a pure environement is fed with a pure input state, the output of the channel has the same eigenvalues as the discarded environment (\textit{i.e.}, the traced-over  output mode of the beam-splitter), as a consequence of the Schmidt decomposition of bipartite pure states.
The combination of these two facts implies that the MOE state of a quantum attenuator with transmittance $\eta$ and pure environment $\ket{\psi}$ is also the MOE state of the quantum attenuator with transmittance $1-\eta$ and environement $\ket{\psi}$, with an extra $\pi$ rotation.

 Although we only investigated the case of quantum attenuators in the present section, we expect our observations to extend to the case of quantum amplifiers.

\section{Conclusions}
\label{sec:conclusions}

In view of the growing interest for non-Gaussian quantum channels in quantum information theory, it is a natural goal to explore the communication properties of such channels. In the present work, we have focused on evaluating precisely the classical capacity of a large class of non-Gaussian quantum channels, that is, the number of classical bits that can reliably be transmitted per use of the channel. We have studied the family of quantum channels known as \textit{quantum attenuators} and \textit{quantum amplifiers}, which couple an input bosonic state to an arbitrary bosonic environment through a Gaussian unitary transformation (either a beam splitter or a two-mode squeezer).
These channels implement the quantum analog of the scaled addition of two random variables at the core of classical additive-noise channels, and have received an increasing attention over the last years.

Our main result is to constrain the classical capacity of these non-Gaussian quantum channels into an interval that is related to the capacity of a Gaussian-equivalent channel. Since this result builds on a related result about the capacity of classical non-Gaussian channels \cite{IHARA197834}, we use Section \ref{sec:classical} to lay out this classical result into a formalism that allows us to generalize it. In Section~\ref{Sec:quantum}, we then present the derivation of our quantum result.
We relate any non-Gaussian attenuator (or amplifier) channel $\mathcal{M}$ to a Gaussian channel $\mathcal{M}_G$ where the non-Gaussian environment is replaced by its Gaussian associate with the same covariance matrix. As $\mathcal{M}_G$ is a Gaussian channel, its classical capacity $C(\mathcal{M}_G)$ is exactly known.
We then proceed to lower bound the capacity of the non-Gaussian channel $C(\mathcal{M})$ by the one of the Gaussian channel, \textit{i.e.}, $C(\mathcal{M})\geq C(\mathcal{M}_G)$, and to upper bound it by the Gaussian capacity plus some constant, \textit{i.e.}, $C(\mathcal{M})\leq C(\mathcal{M}_G)+\Delta$.
The width of the capacity interval $\Delta$ depends on the non-Gaussian character of $\mathcal{M}$ and is related to its minimum output entropy.

This naturally leads us to search for the states that minimize the output entropy of quantum attenuators, which we carry out in Section \ref{sec:conjectures}.
We first point out that the symmetries of the environment play an important role in the properties of the channel.
Supported by numerical evidence, we conjecture that the output entropy of quantum attenuators with an environment that is phase-invariant (\textit{i.e.}, diagonal in the Fock basis) is minimized by coherent states.
The only caveat is that coherent states may, in some cases, not be the unique minimizers.
Beyond this case of phase-invariant environments, we highlight that the output entropy of attenuators with arbirary environments is sometimes minimized by non-Gaussian states.
Yet, symmetry seems to play a role here and we make the further conjecture that states minimizing the output entropy of a quantum attenuator share the same symmetries as the environment (up to a displacement).
Finally, we observe that, for a fixed non-Gaussian environment, the minimum-output-entropy states may vary as a function of the transmittance of the attenuator.

Directions for future research notably include further work on the search for the minimum-output-entropy states of generalized quantum attenuators.
It is reasonable to anticipate that, for some particular non-Gaussian environments, the states minimizing the output entropy may be determined exactly.
Along this line, proving the conjectures that we laid in Section \ref{sec:conjectures} would be a great step forward.

Finally, it would be interesting to investigate the multimode version of the non-Gaussian quantum channels analyzed here.
In particular, it would be important determining whether the capacity is additive or not for these non-Gaussian channels (as it is the case for Gaussian attenuator and amplifier channels).

\section*{Acknowledgements}
\noindent 
N.J.C. is grateful to the James C. Wyant College of Optical Sciences for hospitality during his sabbatical leave in the autumn 2022, when most of this work was achieved. 
ZVH and SG acknowledge funding support from the Army Research Office (ARO) MURI Program Project on Quantum Network Science, ``Theory and Engineering of Large-Scale Distributed Entanglement'', awarded under grant number W911NF2110325.
N.J.C. acknowledges support from Fonds de la Recherche Scientifique - FNRS under Grant No. T.0224.18 and from the European Union under project ShoQC
within ERA-NET Cofund in Quantum Technologies (Quant-ERA) program.

\bibliography{nongaussianchannel}

\clearpage
\newpage

\appendix

\section{Minimization routine}
\label{apd:min_routine}

We present hereafter a simple routine used to minimize a function of a pure state $f(\ket{\psi})$.
In our context, the function $f$ is the von Neumann entropy of a channel $\mathcal{M}$ acting on an input state $\ket{\psi}$, \textit{i.e.}, $f(\ket{\psi})=S\big(\mathcal{M}(\ket{\psi}\bra{\psi})\big)$.

\subsection*{Parameters}
\begin{align*}
    N_{\mathrm{Fock}}&:\text{cutoff photon-number}
    \\
    N_{\mathrm{init}}&:\text{number of random states generated at initialization}
    \\
    N_{\mathrm{loop}}&:\text{number of loops}
    \\
    N_{\mathrm{it}}&:\text{number of iterations per loop}
    \\
    \delta&:\text{initial deviation}
\end{align*}

\subsection*{Pure state generation}

Draw a random unitary $(N_{\mathrm{Fock}}+1)\times (N_{\mathrm{Fock}}+1)$ matrix $U$ according to the Haar measure.
Then the random pure state is:
\begin{align*}
    \ket{\psi_{\mathrm{rand}}}
    =
    \sum\limits_{n=0}^{N_{\mathrm{Fock}}}
    U_{0n}\ket{n}.
\end{align*}

\subsection*{Routine}

Generate $N_\mathrm{init}$ random pure states.
Define $\ket{\psi_{\mathrm{min}}}$ as $\ket{\psi_{\mathrm{min}}}=\mathrm{argmin}\ f(\ket{\psi})$.

\noindent
Repeat $\lbrace 1\rightarrow 2\rbrace$ $N_{\mathrm{loop}}$ times:
\begin{enumerate}
\item Repeat $\lbrace\mathrm{a}\rightarrow\mathrm{b}\rightarrow\mathrm{c}\rbrace$ $N_{\mathrm{it}}$ times:
\begin{enumerate}
    \item
    Generate a random $\ket{\psi_{\mathrm{rand}}}$.
    \item $\ket{\psi_{\mathrm{test}}}=\ket{\psi_{\mathrm{min}}}+\delta\ket{\psi_{\mathrm{rand}}}$. Normalize $\ket{\psi_{\mathrm{test}}}$.
    \item 
    If $f(\ket{\psi_{\mathrm{test}}})<f(\ket{\psi_{\mathrm{min}}}$, then $\ket{\psi_{\mathrm{min}}}\leftarrow\ket{\psi_{\mathrm{test}}}$.
\end{enumerate}
\item Reduce $\delta$: $\delta\leftarrow\delta/2$
\end{enumerate}

\end{document}